# Misrepair mechanism: a mechanism essential for individual adaptation, species' adaptation and species' evolution

Jicun Wang-Michelitsch[1]*, Thomas M Michelitsch[2]

[1]Department of Medicine, Addenbrooke's Hospital, University of Cambridge, UK (Work address until 2007)

[2]Institut Jean le Rond d'Alembert (Paris 6), CNRS UMR 7190 Paris, France

## Abstract

In Misrepair-accumulation theory, we have proposed a Misrepair mechanism for interpreting aging. However, Misrepair mechanism is also important in biological adaptation. Misrepair is a strategy of repair for increasing the surviving chance of an organism when it suffers from severe injuries. As a surviving strategy, Misrepair mechanism plays also an important role in individual adaptation, species' survival, and species' differentiation. **Firstly**, Misrepair of an injury is one of the manners of individual adaptation; and Misrepair mechanism gives an organism a great potential in adapting to changeable and destructive environment. **Secondly**, Misrepair mechanism is important in maintaining and enlarging the diversity of genome DNAs of a species; and a large diversity of genome DNAs is essential for the adaptation and the differentiation of a species in different environments. On one hand, somatic Misrepairs are essential for maintaining the sufficient number of individuals in a species, which are the vectors of different genome DNAs. On the other hand, Misrepair of DNA is a source of DNA mutations and the DNA Misrepairs in germ cells may contribute to the diversity of genome DNAs in a species. In conclusion, Misrepair mechanism is a mechanism essential for individual adaptation, species' adaptation, and species' evolution.

Adaptation to environment for survival is a universal phenomenon in biological world. Individual adaptation is the phenomenon that an organism is able to make suitable responses to the changes in environment for survival. Species' adaptation is the phenomenon that all individuals of a species have special properties to survive in certain environments. Species' stability and the increase of number of individuals are the results of species' adaptation. A large diversity of genome DNAs is the basis for species' adaptations and for species' differentiation in different environments. In Misrepair-accumulation theory, we have proposed a Misrepair mechanism for explaining aging: aging is a process of accumulation of Misrepairs (Wang, 2009). Misrepair mechanism is essential for the survival of a species; and aging of individuals is a sacrifice. In fact, Misrepair mechanism is also essential for maintaining the diversity of genome DNAs in a species. In the present paper, we will discuss the importance of Misrepair mechanism in individual adaptation, species' adaptation, and species' differentiation. Our discussion tackles the following issues:

I. A generalized concept of Misrepair

II. Somatic Misrepair: a strategy for individual adaptation

III. Misrepair mechanism in species' adaptation and in species' evolution

    3.1 Somatic Misrepairs: essential for maintaining the diversity of DNAs of a species

    3.2 DNA Misrepairs in germ cells: contributing to the diversity of genome DNAs

IV. Conclusions

## I. A generalized concept of Misrepair

For explaining aging changes, we have proposed a generalized concept of Misrepair in our novel aging theory, the Misrepair-accumulation theory (Wang, 2009). The term of Misrepair in this theory is different from that in the "Misrepair of DNA". Our new concept of Misrepair is defined as ***an incorrect reconstruction of an injured living structure.*** It is applicable to all living structures including molecules (DNAs), cells, tissues, and organs. For severe injuries occurred to an organism, when a complete repair is impossible to achieve, Misrepair is a way of repair that is essential for maintaining the structural integrity for preventing the death of an organism. It is important to distinguish three concepts: damage, injury and Misrepair. For avoiding confusions, in our theory, the term of damage is referred to an overload to a structure, which may result in an injury of the structure. An injury is a defect of a structure before repair. Misrepair is an incorrect reconstruction of an injured structure, resulting in an alternation of the structure (Figure 1). In our discussion, the term of Misrepair is referred to not only the process but also the result of Misrepair, namely the alteration of a structure after repair.

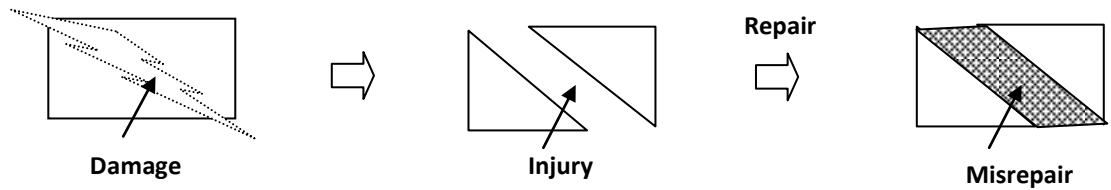

**Figure 1.** The relationship between damage, injury and Misrepair

For understanding aging, it is important to distinguish three concepts: damage, injury and Misrepair. For avoiding confusions, in Misrepair-accumulation theory, the term of damage is referred to an overload to a structure, which may result in an injury of the structure (**Damage**). An injury is a defect of a structure before repair (**Injury**). Misrepair is an incorrect reconstruction of an injured structure, resulting in an alternation of the structure (**Misrepair**).

Misrepairs are irreversible and irremovable; therefore they accumulate and deform gradually a living structure, appearing as aging of the structure. Aging of an organism is therefore a process of accumulation of Misrepairs. Without Misrepairs, an individual could not survive till the age of reproduction; thus Misrepair mechanism is essential for the survival of a species. Aging of individuals is a sacrifice for species' survival. Being beneficial for species' survival is the evolutionary advantage of aging mechanism. In an organism, Misrepairs have a tendency to accumulate to the area of a tissue where an old Misrepair has taken place, because this part of tissue has increased damage-sensitivity and reduced repair-efficiency. Accumulation of Misrepairs is thus self-accelerating and focalized. The process of aging is self-accelerating, and the distribution of aging changes is inhomogeneous. For example, the developments of age spots in the skin and atherosclerotic plaques in arterial walls are both self-accelerating and inhomogeneous.

## II. Somatic Misrepair: a strategy for individual adaptation

A living organism is able to maintain its internal stability and adapt to changeable environments. A change in environment exerts effects to an organism as a stress or as an injury. There are three manners of responses: stress response, complete repair of an injury, and Misrepair of an injury. When an environment change does not cause injuries to an organism, it functions as a stress. Cells and tissues can make suitable responses to the stress to avoid death of cells, and this is called stress response or irritability. Stress response is a physiological process, and the changes in cells/tissues will be restored when the stress is withdrawn. For example, in hot weather, perspiration of animals is a way to maintain the body temperature. The perspiration will be stopped when the environment temperature is reduced. An injury of a structure is a defect of a cell or a tissue. Complete repair takes place when the defect is small, but Misrepair has to be promoted when the defect is too large. In the adaptation by stress response and that by complete repair, the change of a cell/tissue is reversible. Differently, in the adaptation by Misrepair, the change is irreversible (Figure 2). In destructive environments, Misrepair mechanism gives a living being a great potential of environment-adaptation.

**A. Adaptation by stress response**

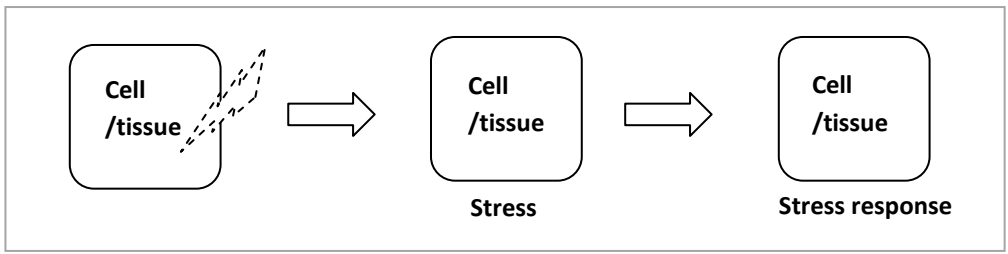

**B. Adaptation by complete repair**

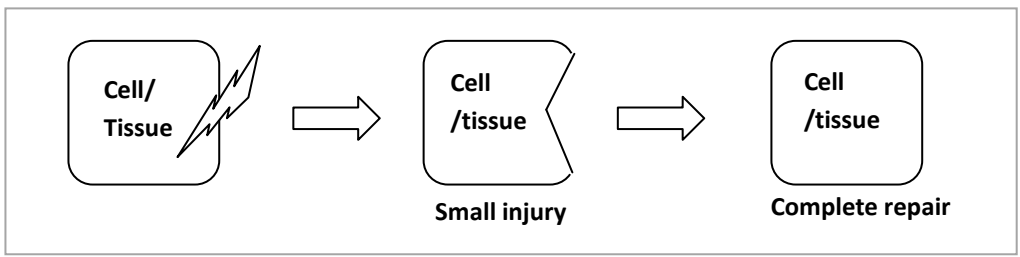

**C. Adaptation by Misrepair**

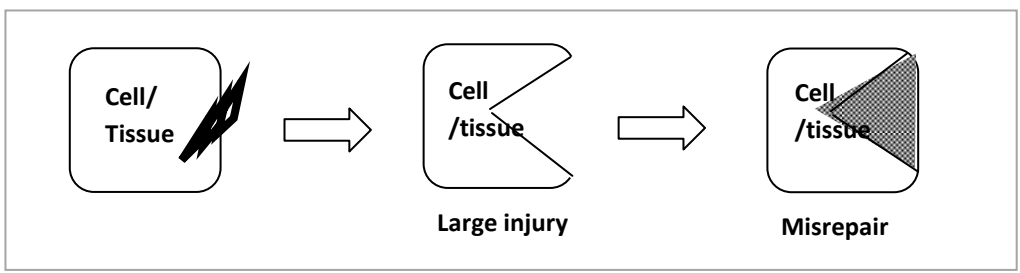

**Figure 2. Three manners of individual adaption: stress response, complete repair and Misrepair**

A change in environment ( ) exerts effects to an organism as a stress or as an injury. An organism has three manners of responses: stress response, complete repair of an injury, and Misrepair of an injury. Stress response is a physiological process, and the changes in a cell/tissue will be restored when the stress is withdrawn (**A**). Complete repair takes place when an injury is small (**B**), but Misrepair has to be promoted when an injury is too large (**C**). In the adaptation by stress response and that by complete repair, the change of a cell/tissue is reversible (**A** and **B**); whereas in the adaptation by Misrepair, the change is irreversible (**C**).

For example, scar formation is a result of Misrepair, which is essential and irreversible. Scar formation is an adaptive response, and it makes the local skin more resistant to chemical and physical damage. Esophagus intestinal metaplasia is also an adaptive change by Misrepair. Replacement of part of the squamous epithelium of distal esophagus by metaplastic glandular epithelium can increase the resistance of the esophagus mucosa to some chemical damage. The myofibers in arterial walls have low potential of regeneration. In arteriosclerosis, the myofibers have two kinds of responses to stress and injuries: one is enlargement of myofibers and the other is proliferation of myofibers. In our view, these two changes are promoted by different degrees of over-loading to myofibers by the movements of contractions/dilatations of arterial walls. The enlargement of myofibers is promoted by an endurable load (stress) or a small injury for functional compensation. Differently, proliferation of myofibers is promoted by death of a myofiber cell.

Adaptation is for survival; however Misrepair will result in irreversible change of the structure of a cell/tissue, leading to aging of it. Adaptation delays the death of an individual and enables the individual to survive till reproduction age. By adaptation and by Misrepairs, genome DNAs of the individual can be transmitted to next generation. Although individuals all die finally, but the genome DNAs of the species are maintained in generation by generation. The potential of individual adaptation is built in structural complexity of an organism, and it is finally determined by the gene configuration of the species. The organisms that have higher structural complexity will have higher potential of individual adaptation. Such a species is more stable than that of simpler organisms in changeable environments. Single-cell organisms such as bacteria have low potential of individual adaptation, and the species' of bacteria are not stable.

## III. Misrepair mechanism in species' adaptation and in species' evolution

Existence of a species is a result of survival of its individuals in changeable environments through adaptation. Species' adaptation is marked by the increase of the number of its individuals and the species' stability in certain environments. Survival of a species is based on the individual adaptation and the reproduction of individuals. Misrepair mechanism plays an important role in species' adaptation and species' evolution. On one hand, somatic Misrepairs are essential for maintaining the sufficient number of individuals in a species and for maintaining the diversity of genome DNAs. On the other hand, Misrepairs on DNA in germ cells contribute to the diversity of genome DNAs in a species.

### 3.1 Somatic Misrepairs: essential for maintaining the diversity of DNAs in a species

A large diversity of DNAs is important for a species on two aspects: **A.** it is beneficial for the survival of a species in changeable environments, and **B.** it provides the substantial basis for species' differentiation. Individuals of a species are the carriers of genome DNAs, and a large diversity of DNAs needs a large number of carriers. These DNA-carriers have similarity and difference to each other on properties, by which they are able to survive in different environments. For multi-cellular beings, somatic Misrepairs reduce the risk of death of an organism (a DNA-carrier) before reproduction age; therefore they are important in maintaining the sufficient number of DNA-carriers in the species. In another word, the Misrepair mechanism in multi-cellular organisms plays an important role in maintaining the stability of a species.

In an isolated environment, when some individuals with certain genetic backgrounds and adaptive properties have higher chance than others for survival and reproduction, they will gradually develop into a sub-group of the species, and this leads to the differentiation of the species. Differentiation of skin color makes the species of human being be able to survive in different climates; and a large diversity of color genes is the substantial basis. In nature, a species with larger diversity of DNAs and bigger population of individuals has higher chance than other species to survive and differentiate.

### 3.2 DNA Misrepairs in germ cells: contributing to the diversity of genome DNAs

Enlargement of the diversity of genome DNAs in a species is made by reproduction. There are two ways of reproduction: sexual reproduction and asexual reproduction. Asexual reproduction is often the way of reproduction for simple organisms including single-cell organisms such as bacterial. In such organisms, the "somatic" cells are also germ cells; and the reproduction is made by the regeneration of cells. Therefore, the DNA changes occurred in all cells can be transmitted to next generations, and they contribute to the diversity of genome DNAs in the species. DNA changes have two types: chromosome changes and point DNA mutations (called DNA mutations). A chromosome change on number or on structure can affect multiple genes, thus it is often fatal to a cell. Differently, point DNA mutations can be mild or silent, and they can survive and accumulate in cells. Therefore, DNA mutations are the main type of DNA changes that can be transmitted by reproduction.

Misrepair of DNA is the main source of DNA mutations in somatic cells (Suh, 2006; Natarajan, 1993; Bishay, 2001). A DNA injury (break) is the promoter of generation of a DNA mutation in a cell (Kasparek, 2011). DNA breaks can be made by radiation, DNA-damaging chemicals, or viral attack. When a DNA break is small, it can be fully repaired. However, when double strands of a DNA disrupt, no template DNA can be used for re-linking the DNA correctly. In this case, altered repair pathways such as non-homologous end joining of broken DNAs have to be promoted for re-linking the DNA. Otherwise the cell will die from failure of DNA (Moore, 1996). Such a repair may result in alteration of local DNA sequence, such as alternation, deletion, or insertion of one or two bases. Therefore the repair in this way is a "Misrepair". Misrepair of DNA is an essential process for transforming a DNA break into a "survivable and inheritable" DNA mutation (Wang-Michelitsch, 2015). Integration of viral DNA into the host DNA such as bacterial DNA is also a result of Misrepair of DNA. The insertion of viral DNA into host DNA can be only achieved by the repair system of the host cell including the DNA lingase. Therefore, for the single-cell species', DNA Misrepairs contribute to the diversity of genome DNAs.

Differently, for the organisms by sexual reproduction, germ cells are produced in reproduction organs. The DNA changes occurred in somatic cells in other organs cannot affect germ cells. Only those DNA changes occurred in germ cells can be transmitted to next generations. Since DNA mutations can accumulate in cells by cell regeneration, germ cells will have increased risk of DNA mutations with age of the organism. The silent DNA mutations in germ cells will be transmitted generation by generation, and become part of the reservoir of genome DNAs of the species. In sexual reproduction, the diversity of genome DNAs is enlarged mainly by non-homologous DNA recombination by fusion of two germ cells that have different genetic backgrounds. The exchange of sister chromatids during meiosis by homologous recombination increases additionally the diversity of DNAs in germ cells and in the species.

Single-cell organisms have high fertility, but they have as well high frequency of DNA injuries and DNA mutations. These species' are unstable and they evolve fast. In contrast, those species' that have complex body structure and make sexual reproduction, such as human being, have lower fertility than single-cell organisms. But they have as well lower risk of DNA injuries and DNA mutations in germ cells. Thus, the species' with higher structural complexity are more stable and they evolve slower. Nevertheless, no matter it is in a simple organism or in a complex organism, important is that DNAs are maintained and transmitted in a great amount of copies and in a large diversity in each species. Survival of DNAs is the final symbol of survival of lives (Dawkins, 2006).

## IV. Conclusions

In interpreting aging, we have proposed a Misrepair mechanism in the Misrepair-accumulation theory. We found out that Misrepair mechanism as a surviving mechanism plays also an important role in individual adaptation, species' adaptation and species evolution. On one hand, the somatic Misrepairs in individuals are essential for individual adaptation and for maintaining the number of individuals, which are the vectors of different genome DNAs in a species. On the other hand, the DNA Misrepairs in germ cells contribute to the diversity of genome DNAs (Figure 3). Taken together, Misrepair mechanism is important in maintaining and enlarging the diversity of genome DNAs, which is the basis for the adaption and the differentiation of a species in nature.

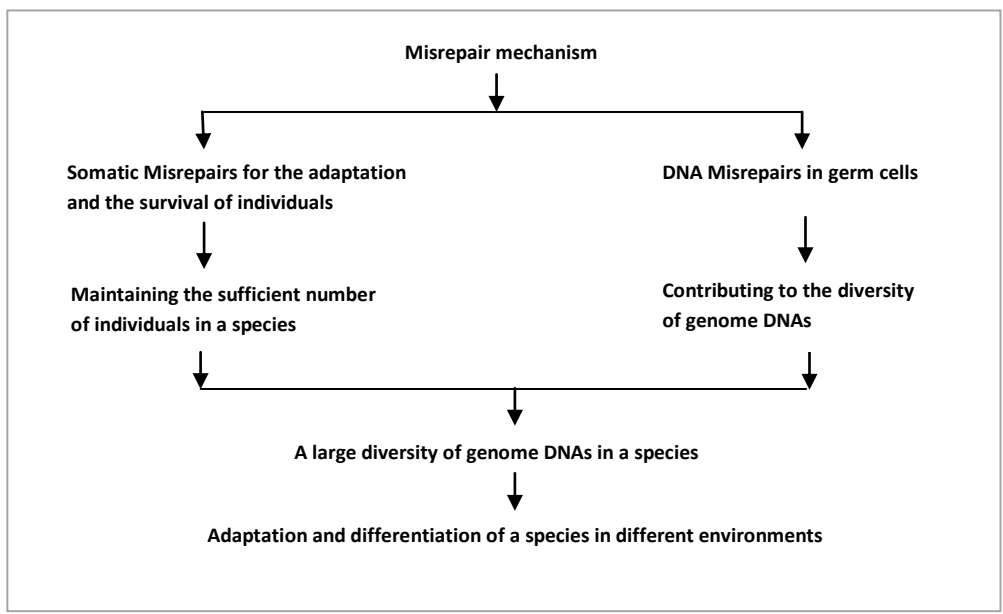

Figure 3. Misrepair mechanism: important in species' adaptation and species' evolution

Misrepair mechanism, as a surviving mechanism, plays an important role in individual adaptation, species' adaptation and species evolution. On one hand, the somatic Misrepairs on individuals are essential for maintaining the sufficient number of individuals, which are the vectors of different genome DNAs in a species. On the other hand, the DNA Misrepairs in germ cells contribute to the diversity of genome DNAs. Taken together, Misrepair mechanism is important in maintaining the diversity of genome DNAs, which is the basis for the adaption and the differentiation of a species in different environments.

In conclusion, Misrepair-accumulation theory helps us understand not only aging and aging-related diseases, but also biological adaptation and species' evolution. In this aspect, Misrepair-accumulation theory is an important support and supplementary to Darwin's Evolution Theory.